\documentclass[12pt]{article}
\usepackage{amsmath,amssymb,amsthm,amsxtra,overpic,bm,epsfig,subfigure}
\usepackage{color}
\usepackage{comment}
\usepackage[all,cmtip]{xy}
\usepackage{cite}
\usepackage[colorlinks,linkcolor=blue]{hyperref}
\usepackage{multirow}

\usepackage{url}
\usepackage{slashed,stmaryrd}
\usepackage{ulem}

\renewcommand{\thefootnote}{\fnsymbol{footnote}}

\def\slash{\!\!\!/}

\begin{document}

\vspace{0.2cm}

\begin{center}
{\large\bf Heavy Majorana Neutrino Production at Future $ep$ Colliders}
\end{center}

\vspace{0.2cm}

\begin{center}
{\bf Shi-Yuan Li} \footnote{E-mail: lishy@sdu.edu.cn}, \quad {\bf Zong-Guo Si} \footnote{E-mail: zgsi@sdu.edu.cn}, \quad {\bf Xing-Hua Yang} \footnote{E-mail: yangxh@mail.sdu.edu.cn}
\\
{School of Physics, Shandong University, Jinan, Shandong 250100, China}\\

\end{center}

\vspace{1.5cm}

\begin{abstract}
The heavy singlet Majorana neutrinos are introduced to generate the neutrino mass in the so-called phenomenological type-I seesaw mechanism. The phenomena induced by the heavy Majorana neutrinos are important to search for new physics. In this paper, we explore the heavy Majorana neutrino production and decay at future $e^{-}p$ colliders. The corresponding cross sections via $W$ and photon fusion are predicted for different collider energies. Combined with the results of the heavy Majorana neutrino production via single $W$ exchange, this work can provide helpful information to search for heavy Majorana neutrinos at future $e^{-}p$ colliders.
\end{abstract}

\begin{flushleft}
\hspace{0.88cm} {\bf Keywords}:  type-I seesaw, Majorana neutrino, $e^{-}p$ collision
\end{flushleft}

\begin{flushleft}
\hspace{0.88cm} PACS number(s):  14.60.St, 13.85.Qk
\end{flushleft}

\def\thefootnote{\arabic{footnote}}
\setcounter{footnote}{0}

\newpage

\section{Introduction}\label{sec1}

The neutrino oscillation experiments show that the neutrinos have minor non-zero masses, which are compelling for new physics beyond the standard model. The phenomena induced by the neutrino mass generating mechanism are important to search for new physics and attract more and more attention. To generate the neutrino mass, the natural way is to introduce the seesaw mechanism. Among them, one simple model is to introduce the heavy right-handed Majorana neutrinos $N_{\rm R}$, which is known as the famous type-I seesaw mechanism~\cite{Minkowski:1977sc,Yanagida:1979ss,Gell-Mann:1979ss,Glashow:1979ss,Mohapatra:1979ia}. The key point to test the type-I seesaw model is to search for the existence of $N_{\rm R}$. In this kind of model, besides the Dirac neutrino mass term $\overline{\nu_{\rm L}} M_{\rm D} N_{\rm R}$, which comes from the Yukawa interactions, the right-handed neutrino $N_{\rm R}$ and its charged-conjugate counterpart $(N_{\rm R})^c$ can also form a Majorana neutrino mass term $\overline{(N_{\rm R})^c}M_{\rm R} N_{\rm R}$, where the effective neutrino mass matrix can be given by the seesaw formula $M^{}_\nu\approx-M_{\rm D}M_{\rm R}^{-1}M_{\rm D}^{\rm T}$, and the smallness of $M^{}_\nu$ can be attributed to the large mass scale of $M_{\rm R}$. Explicitly, the Majorana neutrino mass term $\overline{(N_{\rm R})^c}M_{\rm R} N_{\rm R}$ violates the lepton-number by two units~($\Delta L =2$), so we can probe the Majorana neutrino production signal through the lepton-number violating processes.

The search for Majorana neutrinos via the lepton-number violating process has been studied. At the low Majorana neutrino mass region, the experiments of the interesting processes, such as the well known neutrinoless double beta decays ($N(A,Z) \rightarrow N(A, Z+2) + 2e^-$)~\cite{Furry:1939qr}, the rare decays of the meson ($M_{1}^{\pm} \rightarrow M_{2}^{\mp}\ell_{1}^{\pm}\ell_{2}^{\pm}$)~\cite{Littenberg:1991ek} and tau decays (${\rm e.g.}~\tau^{\pm} \rightarrow \ell^{\mp}M_{1}^{\pm}M_{2}^{\pm}$)~\cite{Ilakovac:1995km} are explored to set the strong constraints on both the heavy Majorana neutrino mass and the related mixing parameters with the charged-leptons~\cite{Atre:2009rg}. For the Majorana neutrino mass above the electroweak scale, its production has been investigated at various collider experiments (for review, we refer to Ref.~\cite{Deppisch:2015qwa,Cai:2017mow}). At hadron colliders, the most widely studied mode is the Drell-Yan process via a single virtual $W$ boson~\cite{Keung:1983uu,Han:2006ip,Perez:2009mu}. Due to the collinear logarithmic enhancement in $t$-channel exchange of massless gauge boson, the vector boson fusion process is important for the higher Majorana neutrino mass~\cite{Datta:1993nm,Dev:2013wba,Alva:2014gxa,Degrande:2016aje}. At higher collider energies, the gluon luminosity grows faster,  so that the heavy Majorana neutrino production via the gluon fusion process becomes interesting~\cite{Willenbrock:1985tj, Ruiz:2017yyf}.
At $e^{+}e^{-}$ colliders, the Majorana neutrino production can be studied via the processes, e.g. $t$-channel $W^\ast$ exchange, $s$-channel $Z^\ast$ exchange~\cite{delAguila:1987nn,Antusch:2016ejd}. Complementary to search for the lepton-number violating processes at hadron colliders and $e^{+}e^{-}$ colliders, in this work, we focus on the heavy Majorana neutrino production and decay in the context of $W^\ast\gamma$ interaction at $e^{-}p$ colliders, which is sub-dominant contribution with respect to the $t$-channel $W^\ast$ exchange process~\cite{Buchmuller:1990vh,Liang:2010gm,Das:2018usr}. These studies are helpful to search for heavy Majorana neutrinos in various rapidity region at future $e^{-}p$ colliders.  As shown in Ref.~\cite{Atre:2009rg}, the general amplitude of $\Delta L =2$ process is proportional to the Majorana neutrino mass, and the light Majorana neutrino contribution is strongly suppressed due to the small neutrino mass, we only consider the heavy Majorana neutrino. The mixing parameter between electron and the heavy Majorana neutrino is strictly constrained~\cite{Belanger:1995nh,Benes:2005hn} and the tau lepton is hard to be reconstructed, therefore we purely concentrate on the di-muon production channel at future $e^{-}p$ colliders, e.g. LHeC~\cite{AbelleiraFernandez:2012cc}, FCC-ep~\cite{Bruning:2260408}, ILC$\otimes$FCC~\cite{Acar:2016rde}.

This paper is organized as follows.  A simple  model is briefly introduced in Section~\ref{sec2},  and  the numerical results and discussions are obtained in Section~\ref{sec3}. Finally, a short summary is given.

\section{Phenomenological type-I seesaw model}\label{sec2}

Within the standard model, neutrinos are massless in the absence of right-handed neutrinos. However, recent neutrino oscillation experiments have clearly shown that neutrinos are massive. In order to explain the smallness of neutrino masses, many new physics models have been proposed. A simple extension of the standard model is to introduce three heavy right-handed neutrino singlets $N_R$ and the gauge-invariant Lagrangian relevant for the neutrino masses can be written as
\begin{eqnarray}
\label{1}
-{\cal L}^{}_{\rm neutrino} = \overline{\ell^{}_{\rm L}} Y^{}_\nu \tilde{H} N^{}_{\rm R} + \frac{1}{2} \overline{(N_{\rm R})^c} M^{}_{\rm R} N^{}_{\rm R} + {\rm h.c.} \; ,
\end{eqnarray}
where $\ell^{}_{\rm L}$ and $\tilde{H} \equiv {\rm i}\sigma^{}_2 H^*$ respectively denote the left-handed lepton doublet and Higgs doublet, and $N^{}_{\rm R}$ the right-handed neutrino singlet. $Y_\nu$ is the $3\times3$ neutrino Yukawa coupling matrix and $M_{\rm R}$ the symmetric right-handed Majorana neutrino mass matrix. After the spontaneous gauge symmetry breaking, the neutrino mass terms appear as
\begin{align}
\label{2}
-{\cal L}_{\rm mass}
&=\overline{\nu_{\rm L}}M_{\rm D}N_{\rm R}+\frac{1}{2} \overline{(N_{\rm R})^c} M^{}_{\rm R} N^{}_{\rm R}+ {\rm h.c.} \nonumber \\
&=\frac{1}{2}\overline{\left(\begin{matrix} \nu_{\rm L} & (N_{\rm R})^c\end{matrix}\right)}
\left(\begin{matrix} 0 & M_{\rm D} \cr M_{\rm D}^{\rm T} & M_{\rm R}\end{matrix}\right)
\left(\begin{matrix} (\nu_{\rm L})^c \cr N_{\rm R} \end{matrix}\right)+ {\rm h.c.} \; .
\end{align}
Here $(\nu_{\rm L})^c$~and $(N_{\rm R})^c$ are respectively defined as $(\nu_{\rm L})^c=C\overline{\nu_{\rm L}}^{\rm T}$~and $(N_{\rm R})^c=C\overline{N_{\rm R}}^{\rm T}$ with $C$ being the charge-conjugation operator. $M_{\rm D} =Y_\nu \langle H\rangle$ is the Dirac neutrino mass matrix with $\langle H\rangle \approx 174~{\rm GeV}$ being the Higgs vacuum expectation value. Since the right-handed neutrinos $N_{\rm R}$ are $SU(2)_{\rm L}$ gauge singlets and thus the Majorana neutrino mass term $\overline{(N_{\rm R})^c}M_{\rm R} N_{\rm R}$ is not subject to the gauge symmetry breaking scale, the absolute scale of the right-handed Majorana neutrino mass matrix $M_{\rm R}$ can naturally be much higher, $M_{\rm R} \gg \langle H\rangle$. In the second line of Eq.~(\ref{2}), we have used the relationship $\overline{(N_{\rm R})^c}M_{\rm D}^{\rm T} (\nu_{\rm L})^c=\overline{\nu_{\rm L}}M_{\rm D}N_{\rm R}$.

The overall mass matrix in Eq.~(\ref{2}) is symmetric and can be diagonalized by one unitary transformation
\begin{eqnarray}
\label{3}
\left(\begin{matrix} V & R \cr S & U \end{matrix}\right)^{\dagger}
\left(\begin{matrix}0 & M_{\rm D} \cr M_{\rm D}^{\rm T} & M_{\rm R}\end{matrix}\right)
\left(\begin{matrix} V & R \cr S & U \end{matrix}\right)^{\ast}
=\left(\begin{matrix} \widehat{M_\nu} & 0 \cr 0 & \widehat{M_N} \end{matrix}\right) \; ,
\end{eqnarray}
where $\widehat{M_\nu}={\rm Diag}\{m_1, m_2, m_3\}$ and $\widehat{M_N}={\rm Diag}\{M_1, M_2, M_3\}$ denote the mass eigenvalues of light and heavy Majorana neutrinos, respectively. In the limit of $M_{\rm D} \ll M_{\rm R}$, the effective neutrino mass matrix can be of the order of $M^{}_\nu\approx-M_{\rm D}M_{\rm R}^{-1}M_{\rm D}^{\rm T}$, and the smallness of $M^{}_\nu$ can be attributed to the largeness of $M_{\rm R}$. According to the unitary condition of the transformation matrix in Eq.~(\ref{3}), the matrices $V, R, S ,U$ can satisfy
\begin{eqnarray}
\label{4}
VV^\dagger+RR^\dagger = SS^\dagger+UU^\dagger =& \textbf{1}\; , \nonumber \\
V^\dagger V+S^\dagger S = R^\dagger R+U^\dagger U =& \textbf{1}\; ,
\end{eqnarray}
with $VV^\dagger \sim UU^\dagger \sim {\cal O}(1)$ and $RR^\dagger \sim S^\dagger S \sim {\cal O}(M_{\rm D}^2/M_{\rm R}^2)$.

Moreover, the relation between the neutrino flavor eigenstates $\nu_\alpha$~(for $\alpha=e, \mu, \tau$) and the mass eigenstates $\nu_i$ and $N_i$~(for $i=1, 2, 3$) can be given by
\begin{eqnarray}
\label{5}
\left(\begin{matrix}\nu_e \cr \nu_\mu \cr \nu_\tau\end{matrix}\right)_{\rm L}=
V\left(\begin{matrix}\nu_1 \cr \nu_2 \cr \nu_3\end{matrix}\right)_{\rm L}+
R\left(\begin{matrix}N_1 \cr N_2 \cr N_3\end{matrix}\right)_{\rm L} \; .
\end{eqnarray}
Therefore the standard weak charged-current interaction Lagrangian of leptons in terms of the mass eigenstates can be written as
\begin{eqnarray}
\label{6}
-{\cal L}_{\rm cc}
=\frac{g}{\sqrt{2}}
\left[\overline{\left(e, \mu, \tau\right)_{\rm L}} \gamma^\mu V \left(\begin{matrix}\nu_1 \cr \nu_2 \cr \nu_3\end{matrix}\right)_{\rm L} W_{\mu}^{-} +
\overline{\left(e, \mu, \tau\right)_{\rm L}} \gamma^\mu R \left(\begin{matrix}N_1 \cr N_2 \cr N_3\end{matrix}\right)_{\rm L} W_{\mu}^{-} \right]+{\rm h.c.} \; .
\end{eqnarray}
It is worth mentioning that we have already chosen the basis where the flavor eigenstates of three charged-leptons are identified with their mass eigenstates.
The matrix $V$ in Eq.~(\ref{6}) is the so-called Pontecorvo-Maki-Nakagawa-Sakata (PMNS) matrix~\cite{Pontecorvo:1957cp,Maki:1962mu}, denotes the mixing between charged-leptons and light Majorana neutrinos and can be measured from the oscillation experiments.
While the matrix $R$ indicates the mixing between charged-leptons and heavy Majorana neutrinos, which can be determined from the $0\nu\beta\beta$-decay experiments or possible collider experiments.

However, within the context of the canonical, high-scale type-I seesaw model, the mass scale of the heavy Majorana neutrino $M_{\rm R}$ is too high to be detected experimentally. Given $M_{\rm D} \sim 10^{2}~{\rm GeV}$, for instance, the scale of $M_{\rm R}$ must be at the order of $10^{13}~{\rm GeV}$ to generate the tiny neutrino masses $M_{\nu} \sim {\rm eV}$. In principle, the scale of the type-I seesaw can be lowered to TeV or below, but this also implies a small mixing $R \sim {\cal O}(M_{\rm D}/M_{\rm R}) \sim {\cal O}(10^{-6})$, which strongly suppresses the heavy neutrino production. In order to increase the mixing $R$, the total contribution to the light neutrino masses have to be cancelled~\cite{Buchmuller:1991tu,Ingelman:1993ve}. Recently, it is shown in Ref.~\cite{Kersten:2007vk,Moffat:2017feq} that if the standard model field content is extended only by singlet fermions, the vanishing light neutrino masses implies that lepton-number is conserved. Hence, if the lepton-number violating processes involving heavy neutrinos are observed at colliders, one can conversely conclude that there is in fact a larger, richer field content than initially assumed.
In this work, in order to investigate the discovery potential of heavy Majorana neutrinos at future $e^-p$ colliders, we do not address how to build a realistic model, but simply use Eq.~(\ref{6}) as a phenomenological Lagrangian.

At present, the constraint on the mixing between heavy Majorana neutrinos and electrons can be derived from the $0\nu\beta\beta$-decay experiments~\cite{Belanger:1995nh,Benes:2005hn}
\begin{eqnarray}
\label{8}
\sum_{N} \frac{|R_{eN}|^2}{m_N} < 5 \times 10^{-5}~{\rm TeV^{-1}} \; , \quad {\rm for}~m_N \gg 1~{\rm GeV} \; .
\end{eqnarray}
For the mixing between heavy Majorana neutrinos and muons, the most stringent bound comes from the LHC experiments~\cite{Sirunyan:2018mtv,Aad:2015xaa}
\begin{eqnarray}
\label{9}
|R_{\mu N}|^2 < 3.2 \times 10^{-3} - 5.0 \times 10^{-2}~({\rm at}~95\%~{\rm C.L.}) \; , \quad {\rm for}~m_N = 100 - 500~{\rm GeV} \; .
\end{eqnarray}
For Majorana neutrinos heavier than the electroweak scale, the mixing between heavy Majorana neutrinos and charged-leptons can be restricted by a global fit to electroweak precision data, tests of CKM unitarity and tests of lepton universality~\cite{Fernandez-Martinez:2016lgt},
\begin{eqnarray}
\label{10}
|R_{e N}|^2 < 2.5 \times 10^{-3}, \quad |R_{\mu N}|^2 < 4.4 \times 10^{-4}, \quad |R_{\tau N}|^2 < 5.6 \times 10^{-3}~({\rm at}~95\%~{\rm C.L.})\; .
\end{eqnarray}

As shown in Ref.~\cite{Atre:2009rg,Si:2008jd}, the total decay width of the heavy Majorana neutrino can be expressed approximately as
\begin{eqnarray}
\label{11}
\Gamma_N \simeq \left\{
  \begin{array}{ll}
  \setlength{\arraycolsep}{20pt}
  \left(\frac{3G_F^2 m_N^5}{32 \pi^3}\right)\sum_{\alpha=e, \mu ,\tau}|R_{\alpha N}|^2 \; , & m_N < m_W \; , \\
  \left(\frac{3G_F m_N^3}{8 \pi \sqrt{2}}\right)\sum_{\alpha=e, \mu ,\tau}|R_{\alpha N}|^2 \; , & m_N > m_W \; .
  \end{array} \right.
  \vspace{-11pt}
\end{eqnarray}

\section{Heavy Majorana neutrino phenomena at $ep$ colliders}\label{sec3}

\begin{figure}[!htbp]
\begin{center}
\subfigure[]{\label{fig1a}
\includegraphics[width=0.40\textwidth]{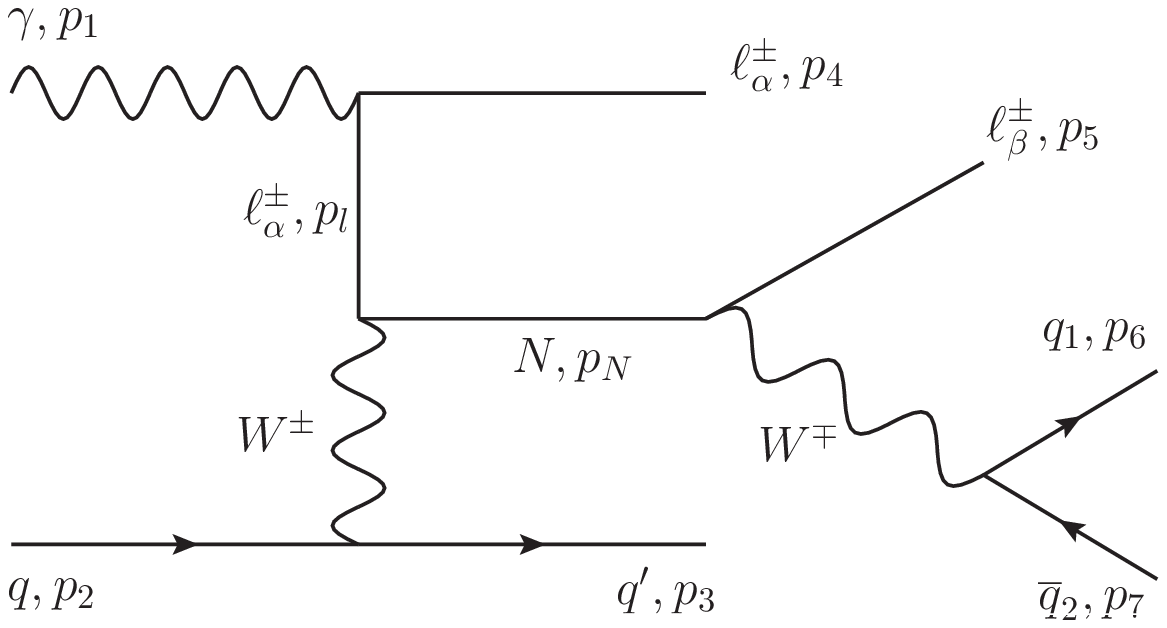} }
\hspace{-1.5cm}~
\subfigure[]{\label{fig1b}
\includegraphics[width=0.50\textwidth]{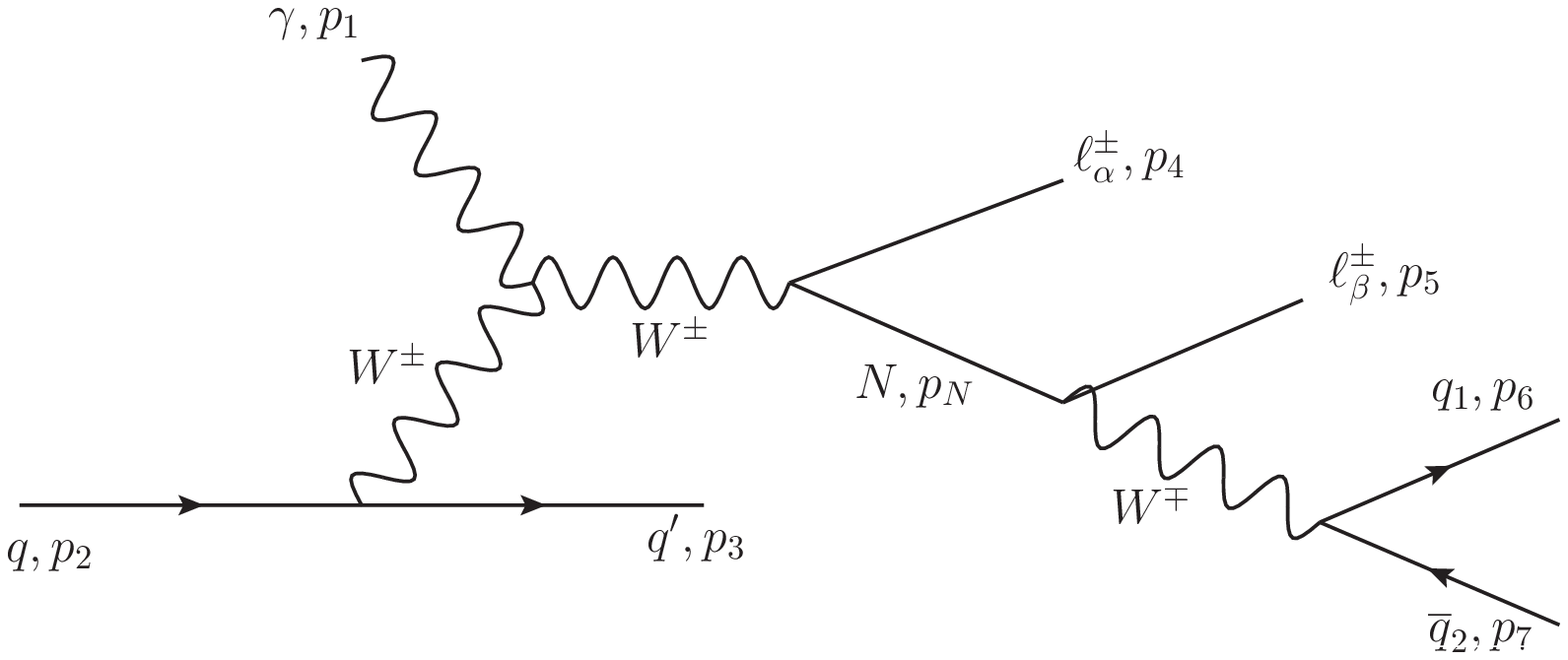} }
\caption{Feynman diagrams at the parton level for the process $\gamma q \rightarrow \ell_\alpha^{\pm}\ell_\beta^{\pm}+X$.}\label{fig1}
\end{center}
\end{figure}

We start by considering the process
\begin{eqnarray}
\label{12}
e^- + p \rightarrow e^- + \ell_\alpha^{\pm} + N + X \rightarrow e^- + \ell_\alpha^{\pm} + \ell_\beta^{\pm} (\ell_\beta^{\mp}) + X  \; .
\end{eqnarray}
The relevant process at the parton level (Fig.~\ref{fig1}) is
\begin{eqnarray}
\label{13}
\gamma(p_1)+q(p_2) \rightarrow q^\prime(p_3) + \ell_\alpha^{\pm}(p_4) + \ell_\beta^{\pm}(\ell_\beta^{\mp})(p_5) +  q_1(p_6) + \overline{q}_2(p_7) \; ,
\end{eqnarray}
where $p_i$ (for $i=1, \cdots, 7$) is the four-momentum of the corresponding particle. The photon is emitted from the electron and can be described by the photon density function~\cite{Frixione:1993yw}
\begin{eqnarray}
\label{14}
f_{\gamma/e^-}(x)=
\frac{\alpha}{2\pi}\left[\frac{1+(1-x)^2}{x}\ln{\frac{Q_{\rm max}^2}{Q_{\rm min}^2}}+ 2 m_e^2 x \left(\frac{1}{Q_{\rm max}^2}-\frac{1}{Q_{\rm min}^2}\right)\right] \; .
\end{eqnarray}
Here $x=E_\gamma/E_e$ with $E_\gamma$ and $E_e$ the energies of the photon and electron, respectively. $\alpha$ is the fine structure constant and $m_e$ the mass of electron. $Q_{\rm min}^2=m_e^2 x^2/(1-x)$ and $Q_{\rm max}^2=(\theta_c E_e)^2(1-x)+Q_{\rm min}^2$ with $\theta_c$ the cut of the electron scattering angle.

The cross section for the process in Eq.~(\ref{12}) can be written as
\begin{eqnarray}
\label{15}
\sigma(e^- p \rightarrow \ell_\alpha \ell_\beta+X)=\sum_{q}\int dx_1 dx_2 f_{\gamma/e^-}(x_1) f_{q/p}(x_2,\mu^2)\cdot \hat{\sigma}(\gamma q \rightarrow \ell_\alpha \ell_\beta+X) \; ,
\end{eqnarray}
where $f_{q/p}(x_2,\mu^2)$ is the parton distribution function with $x_{2}$ the energy fraction of $q$, and $\mu$ the factorization scale. Here, we employ the {\rm CT14QED}~\cite{Schmidt:2015zda} for the photon distribution function and parton distribution functions in proton. $\hat{\sigma}$ is the partonic cross section
\begin{eqnarray}
\label{16}
\hat{\sigma}(\gamma q \rightarrow \ell_\alpha \ell_\beta+X)=\frac{1}{2\hat{s}}\int \overline{|{\cal M}|^2} d{\cal L}ips_5 \; .
\end{eqnarray}
Here $\hat{s}=x_1 x_2 s$ is the flux factor with $\sqrt{s}$ the electron-proton center-of-mass energy. $d{\cal L}ips_5$ represents the five-body Lorentz invariant phase space of the final particles, and $\overline{|{\cal M}|^2}$ is the squared scattering amplitude averaged (summed) over the initial (final) particles.

For the convenience of the discussions on the numerical results, all the input parameters used in our numerical analysis are listed as follows,
\begin{eqnarray}
m_e=0.51~{\rm MeV},~~~~~\theta_{c}=32~{\rm mrad},~~~~~\mu=m_N \; , \nonumber
\end{eqnarray}
\begin{eqnarray}
m_W=80.385~{\rm GeV},~~~~~\Gamma_W=2.085~{\rm GeV},~~~~~G_F=1.166\times10^{-5}~{\rm GeV}^{-2},~~~~~\sin^2{\theta_W}=0.231 \; , \nonumber
\end{eqnarray}
\begin{eqnarray}
\label{17}
|R_{e N}|^2=5.0\times10^{-6},~~~~~ |R_{\mu N}|^2=2.0\times10^{-3},~~~~~ |R_{\tau N}|^2=5.6\times10^{-3} \; .
\end{eqnarray}
For the signal process, we develop a package by the help of Form~\cite{Kuipers:2012rf} to generate a Fortran code, and the numerical integration has been performed with Vegas code~\cite{Lepage:1980dq} in order to obtain a selection of kinematic distributions. The background processes are simulated by MadGraph~\cite{Alwall:2014hca}.

\begin{figure}[!htbp]
\begin{center}
\subfigure[]{\label{fig2a}
\includegraphics[width=0.45\textwidth]{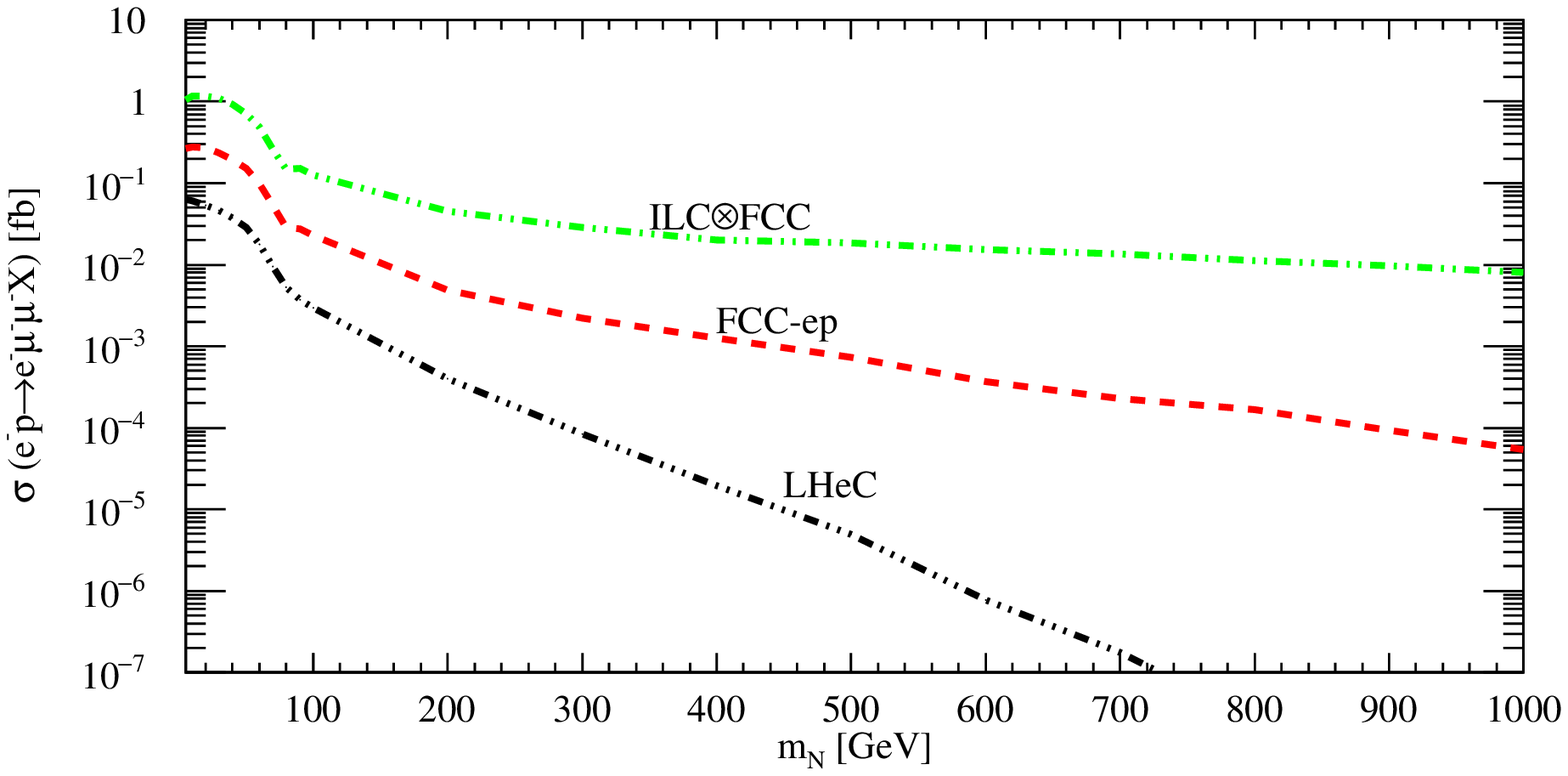} }
\hspace{-1.5cm}~
\subfigure[]{\label{fig2b}
\includegraphics[width=0.45\textwidth]{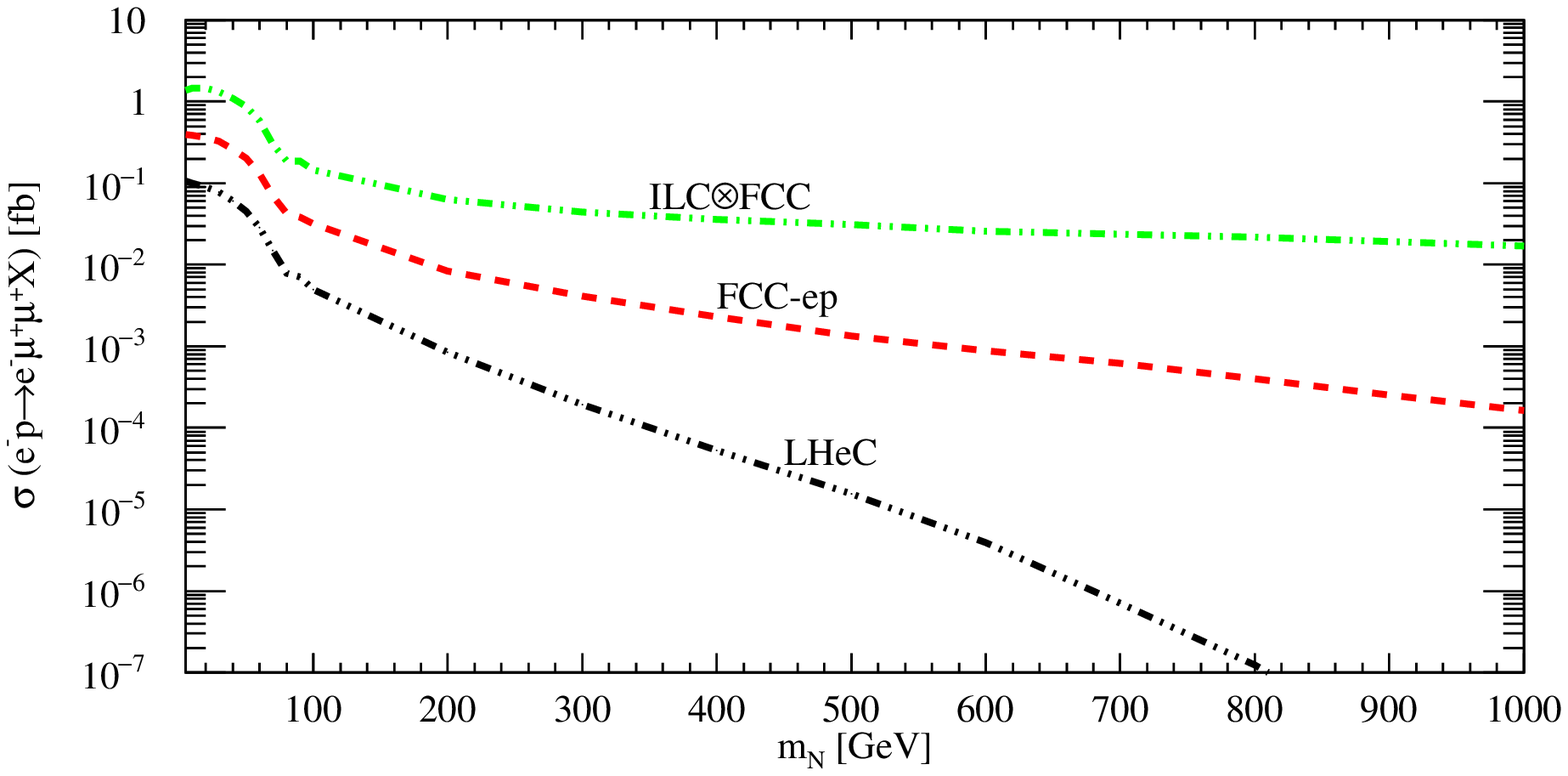} }
\caption{The cross section for (a) $e^-p \rightarrow e^- \mu^{-}\mu^{-}+X$, (b) $e^-p \rightarrow e^- \mu^{+}\mu^{+}+X$ as a function of $m_N$.}\label{fig2}
\end{center}
\end{figure}

In this work, we only consider the contribution of a single heavy Majorana neutrino and concentrate on the di-muon production channel. We obtain the cross section for the process in Eq.~(\ref{12}) at LHeC with $\sqrt{s}=1.3~{\rm TeV}$, FCC-ep with $\sqrt{s}=3.5~{\rm TeV}$ and ${\rm ILC}\otimes{\rm FCC}$ with $\sqrt{s}=10~{\rm TeV}$. The cross sections in final states with same-sign dileptons are shown in Fig.~\ref{fig2} as a function of the heavy Majorana neutrino mass. The difference between $\sigma(e^-p \rightarrow e^- \mu^{+}\mu^{+}+X)$ and $\sigma(e^-p \rightarrow e^- \mu^{-}\mu^{-}+X)$ can be attributed to the role of parton distribution function $f_{q/p}(x,\mu^2)$ and induce the charge asymmetry. To investigate this charge asymmetry, we define
\begin{eqnarray}
\label{18}
{\cal A}_C= \frac{\sigma(e^-p \rightarrow e^- \mu^{+}\mu^{+}+X)-\sigma(e^-p \rightarrow e^- \mu^{-}\mu^{-}+X)}
{\sigma(e^-p \rightarrow e^- \mu^{+}\mu^{+}+X)+\sigma(e^-p \rightarrow e^- \mu^{-}\mu^{-}+X)} \; .
\end{eqnarray}
The numerical results of ${\cal A}_C$ as a function of $m_N$ are listed in Table~\ref{table1}.

\begin{table}[!htbp]
  \centering
  \begin{tabular}{|c|c|c|c|c|c|c|c|}
  \hline
  \hline
  $m_N$~[GeV] & 5 & 10 & 20 & 40 & 60 & 80 & 100 \\
  \hline
  LHeC & 0.245 & 0.240 & 0.246 & 0.228 & 0.236 & 0.199 & 0.257 \\
  \hline
  FCC-ep & 0.187 & 0.169 & 0.149 & 0.144 & 0.142 & 0.191 & 0.158 \\
  \hline
  ILC$\otimes$FCC & 0.127 & 0.108 & 0.107 & 0.085 & 0.089 & 0.106 & 0.066 \\
  \hline
  \hline
  \end{tabular}
  \caption{The charge asymmetry function ${\cal A}_C$ for different heavy Majorana neutrino mass $m_N$ at LHeC, FCC-ep and ILC$\otimes$FCC.}\label{table1}
\end{table}

\begin{figure}[!htbp]
\begin{center}
\includegraphics[width=0.52\textwidth]{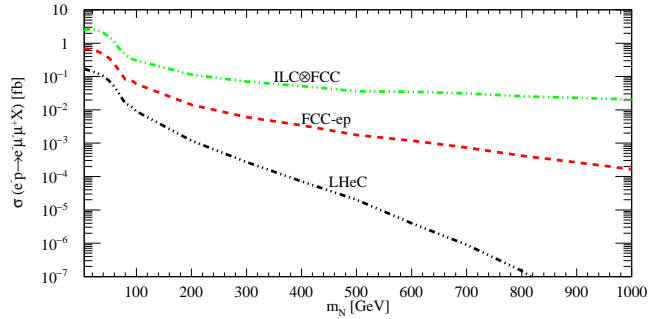}
\caption{The cross section for $e^-p \rightarrow e^- \mu^{-}\mu^{+}+X$ as a function of $m_N$.}\label{fig3}
\end{center}
\end{figure}

We also investigate the process $e^-p \rightarrow e^- \mu^{-}\mu^{+}+X$ of the opposite-sign dileptons, the corresponding cross sections as a function of $m_N$ are displayed in Fig.~\ref{fig3}. For this process, the standard model process for the $\mu^{-}\mu^{+}$ production via a $Z^{0}$ or a virtual photon is the dominant background and can be greatly reduced by the constraint for the invariant mass of the $\mu^{-}\mu^{+}$ pair.

\begin{figure}[!htbp]
\begin{center}
\subfigure[]{\label{fig4a}
\includegraphics[width=0.45\textwidth]{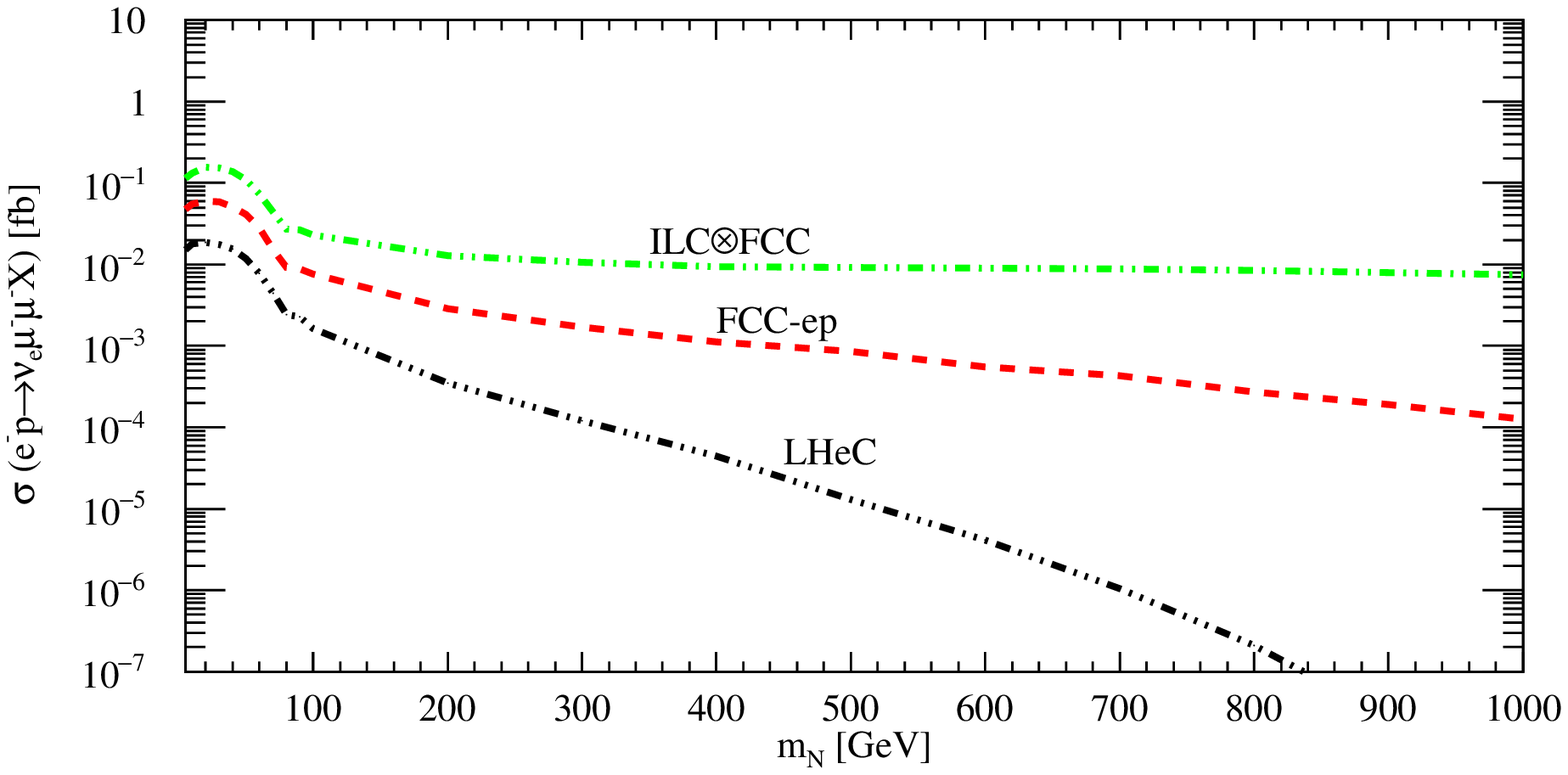} }
\hspace{-1.5cm}~
\subfigure[]{\label{fig4b}
\includegraphics[width=0.45\textwidth]{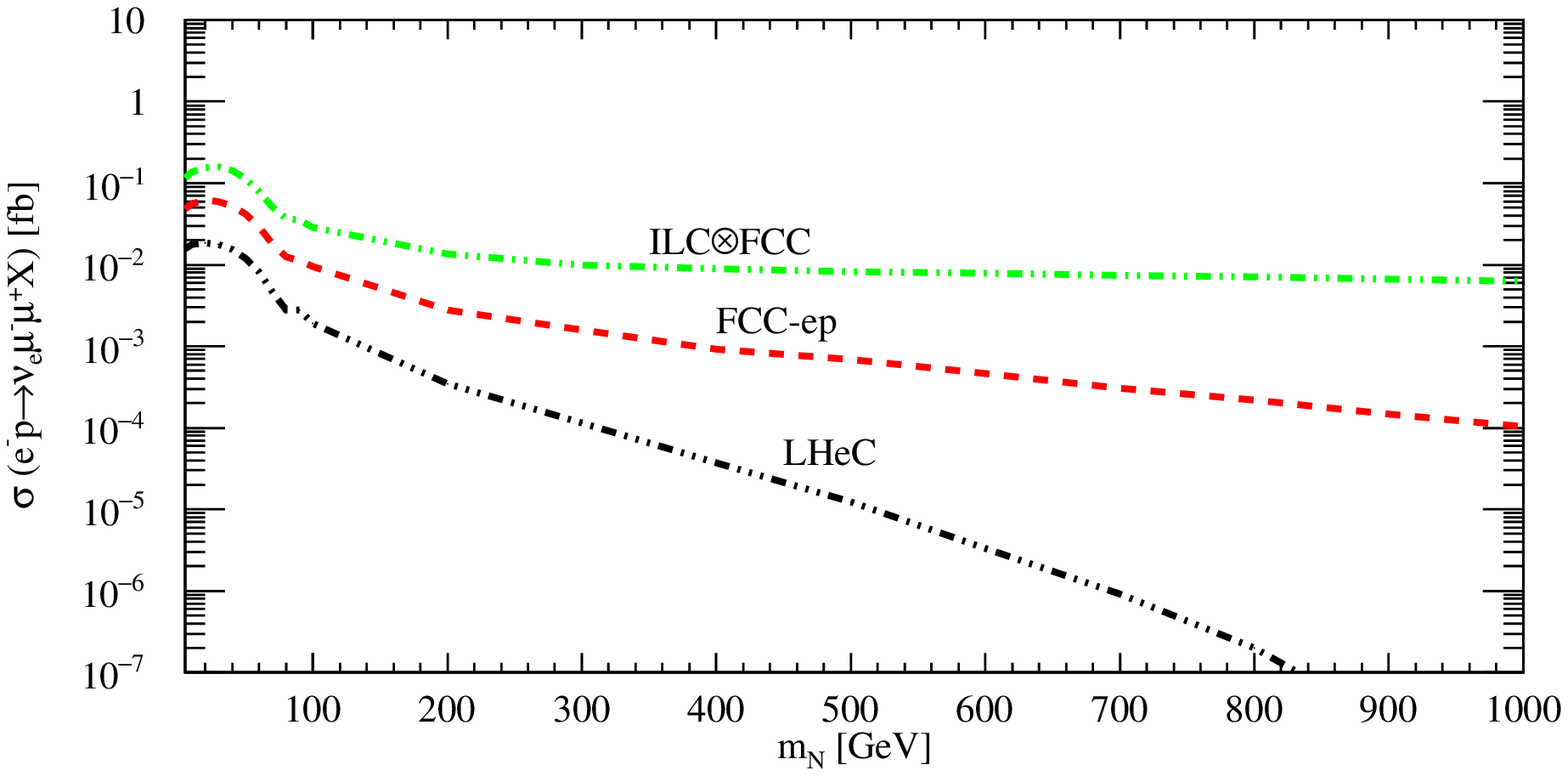} }
\caption{The cross section for (a) $e^-p \rightarrow \nu_e \mu^{-}\mu^{-}+X$, (b) $e^-p \rightarrow \nu_e \mu^{-}\mu^{+}+X$ as a function of $m_N$.}\label{fig4}
\end{center}
\end{figure}

Similar as Eq.~(\ref{12}), we study the process $e^- + p \rightarrow \nu_e + \ell_\alpha^{-} + N + X \rightarrow \nu_e + \ell_\alpha^{-} + \ell_\beta^{\pm} + X$, its cross section can be written as
\begin{eqnarray}
\label{19}
\sigma(e^- p \rightarrow \ell_\alpha \ell_\beta+X)=\int dx_1 f_{\gamma/p}(x_1,\mu^2) \cdot \hat{\sigma}(\gamma e \rightarrow \ell_\alpha \ell_\beta+X) \; ,
\end{eqnarray}
where the photon is emitted from the proton and can be described by the photon distribution function $f_{\gamma/p}(x, \mu^2)$. The results of the corresponding cross sections related to $\mu^-\mu^\pm$ channels are shown in Fig.~\ref{fig4}.
As we employ the ``Improved  Weizsacker-Williams" approximation in Eq.~(\ref{14}) for the photon emitted from the electron,  we are inclusive with respect to final-state $e^-$ below an angle $\theta_c$, and hence cannot distinguish the neutral current splitting $e^- \rightarrow \gamma^\ast e^-$ ($\theta<\theta_c$) from the charged current splitting $e^- \rightarrow W^\ast \nu_e$. After considering the significant impact of the process $e^- p \rightarrow \nu_e \mu^{-} \mu^{-} + X$, we redefine the charge asymmetry as
\begin{eqnarray}
\label{}
{\cal A}^\prime_C= \frac{\sigma(e^-p \rightarrow \mu^{+}\mu^{+}+X)-\sigma(e^-p \rightarrow \mu^{-}\mu^{-}+X)}
{\sigma(e^-p \rightarrow \mu^{+}\mu^{+}+X)+\sigma(e^-p \rightarrow \mu^{-}\mu^{-}+X)} \; ,
\end{eqnarray}
and list the numerical results of ${\cal A}_C^\prime$ in Table~\ref{table2}.

\begin{table}[!htbp]
  \centering
  \begin{tabular}{|c|c|c|c|c|c|c|c|}
  \hline
  \hline
  $m_N$~[GeV] & 5 & 10 & 20 & 40 & 60 & 80 & 100 \\
  \hline
  LHeC & 0.142 & 0.116 & 0.103 & 0.062 & 0.058 & 0.012 & 0.043 \\
  \hline
  FCC-ep & 0.108 & 0.079 & 0.050 & 0.030 & 0.017 & 0.055 & 0.017 \\
  \hline
  ILC$\otimes$FCC & 0.077 & 0.055 & 0.045 & 0.017 & 0.019 & 0.023 & -0.018 \\
  \hline
  \hline
  \end{tabular}
  \caption{The charge asymmetry function ${\cal A}^\prime_C$ for different heavy Majorana neutrino mass $m_N$ at LHeC, FCC-ep and ILC$\otimes$FCC.}\label{table2}
\end{table}

\begin{figure}[!htbp]
\begin{center}
\subfigure[]{\label{fig5a}
\includegraphics[width=0.45\textwidth]{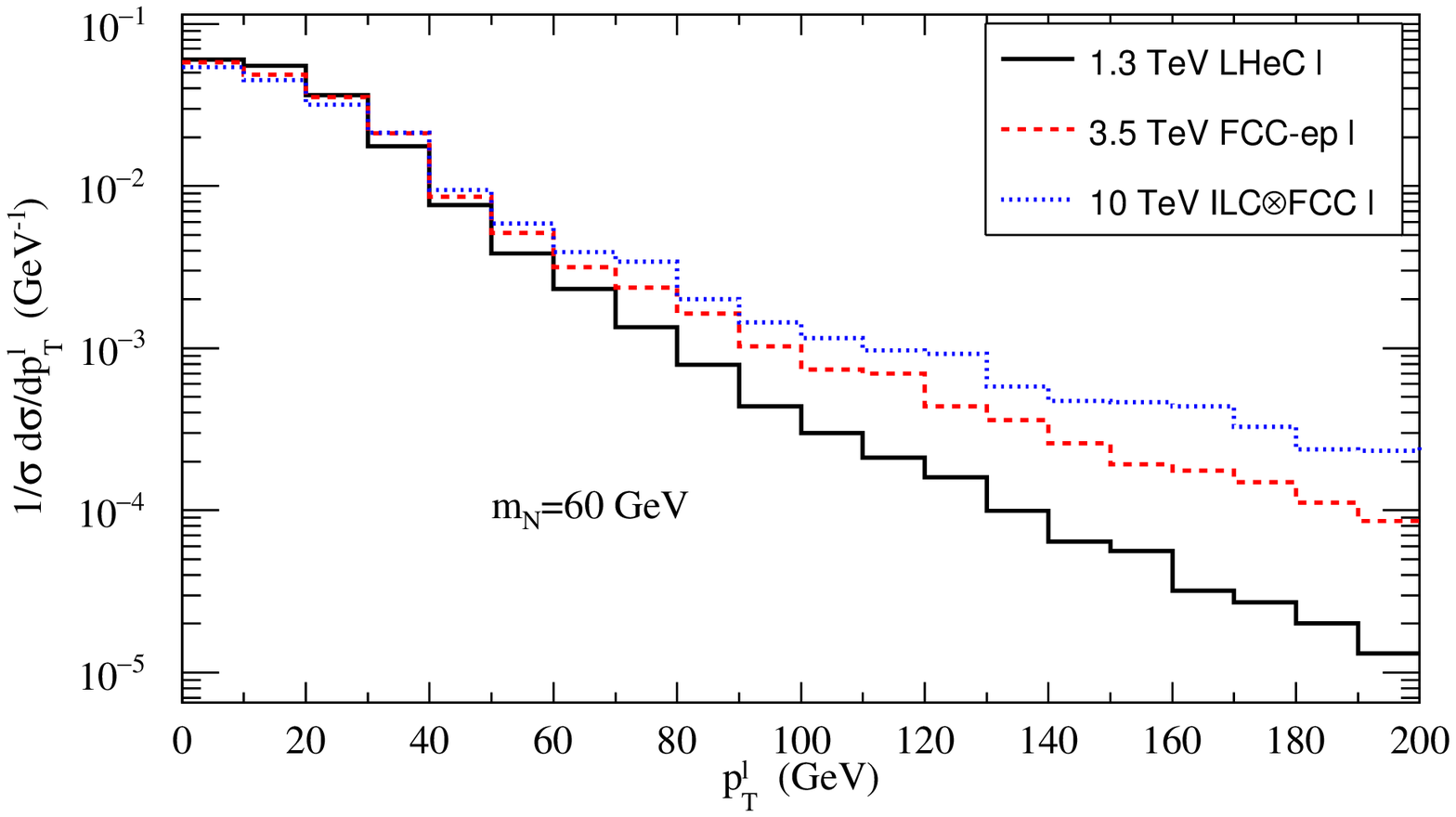} }
\hspace{-1.5cm}~
\subfigure[]{\label{fig5b}
\includegraphics[width=0.45\textwidth]{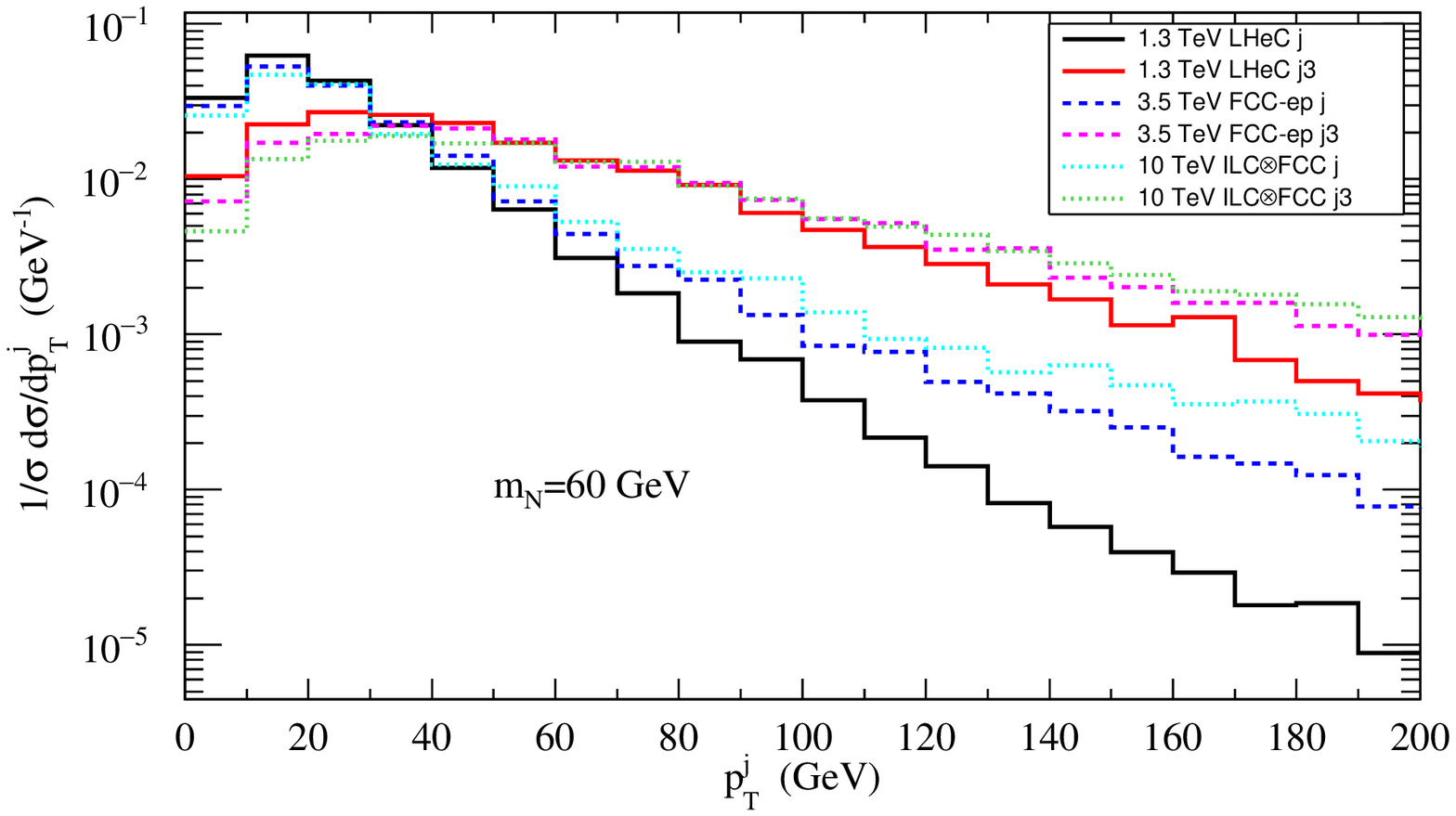} }
\caption{The normalized transverse momentum distributions (a) $1/\sigma{\rm d}\sigma/{\rm d}p^{\ell}_{\rm T}$ and (b) $1/\sigma{\rm d}\sigma/{\rm d}p^{j,j_3}_{\rm T}$ for $m_N=60~{\rm GeV}$ at 1.3~TeV LHeC~(solid), 3.5~TeV FCC-ep~(dash), 10~TeV ILC$\otimes$FCC~(dot).}\label{fig5}
\end{center}
\end{figure}

\begin{figure}[!htbp]
\begin{center}
\includegraphics[width=0.52\textwidth]{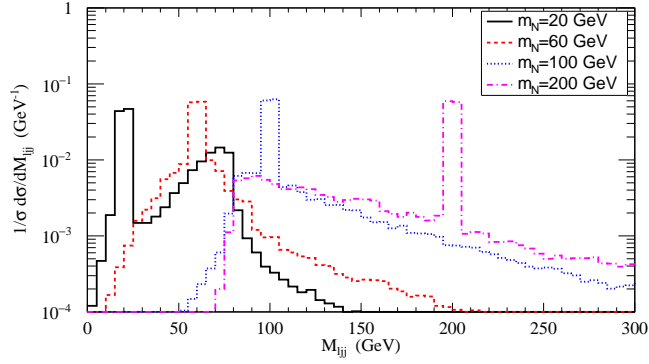}
\caption{The normalized invariant mass distributions $1/\sigma{\rm d}\sigma/{\rm d}M_{\ell jj}$ for $m_N$ = 20, 60, 100, 200 GeV at 3.5~TeV FCC-ep.}\label{fig6}
\end{center}
\end{figure}

\begin{figure}[!htbp]
\begin{center}
\subfigure[]{\label{fig7a}
\includegraphics[width=0.45\textwidth]{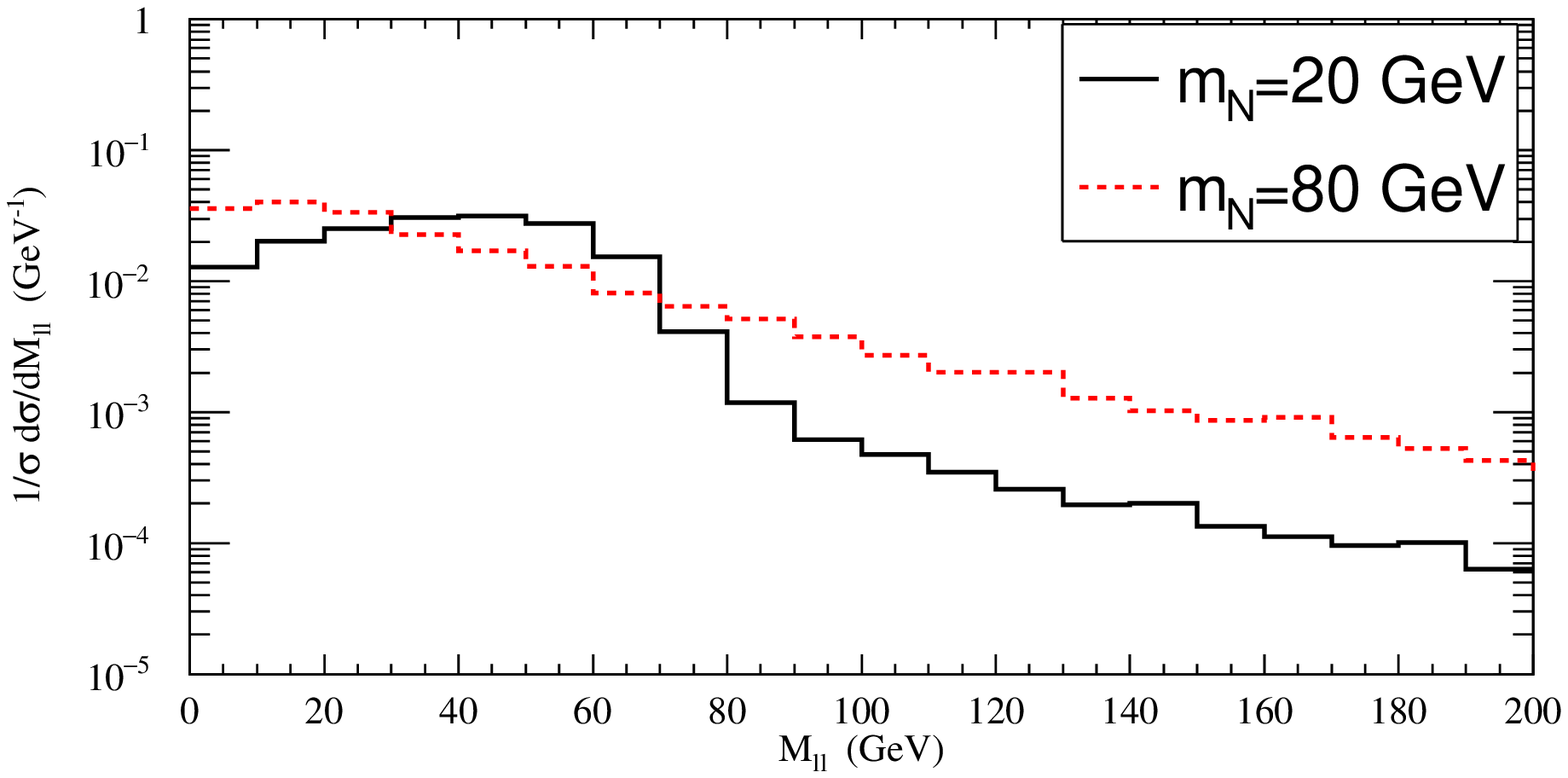} }
\hspace{-1.5cm}~
\subfigure[]{\label{fig7b}
\includegraphics[width=0.45\textwidth]{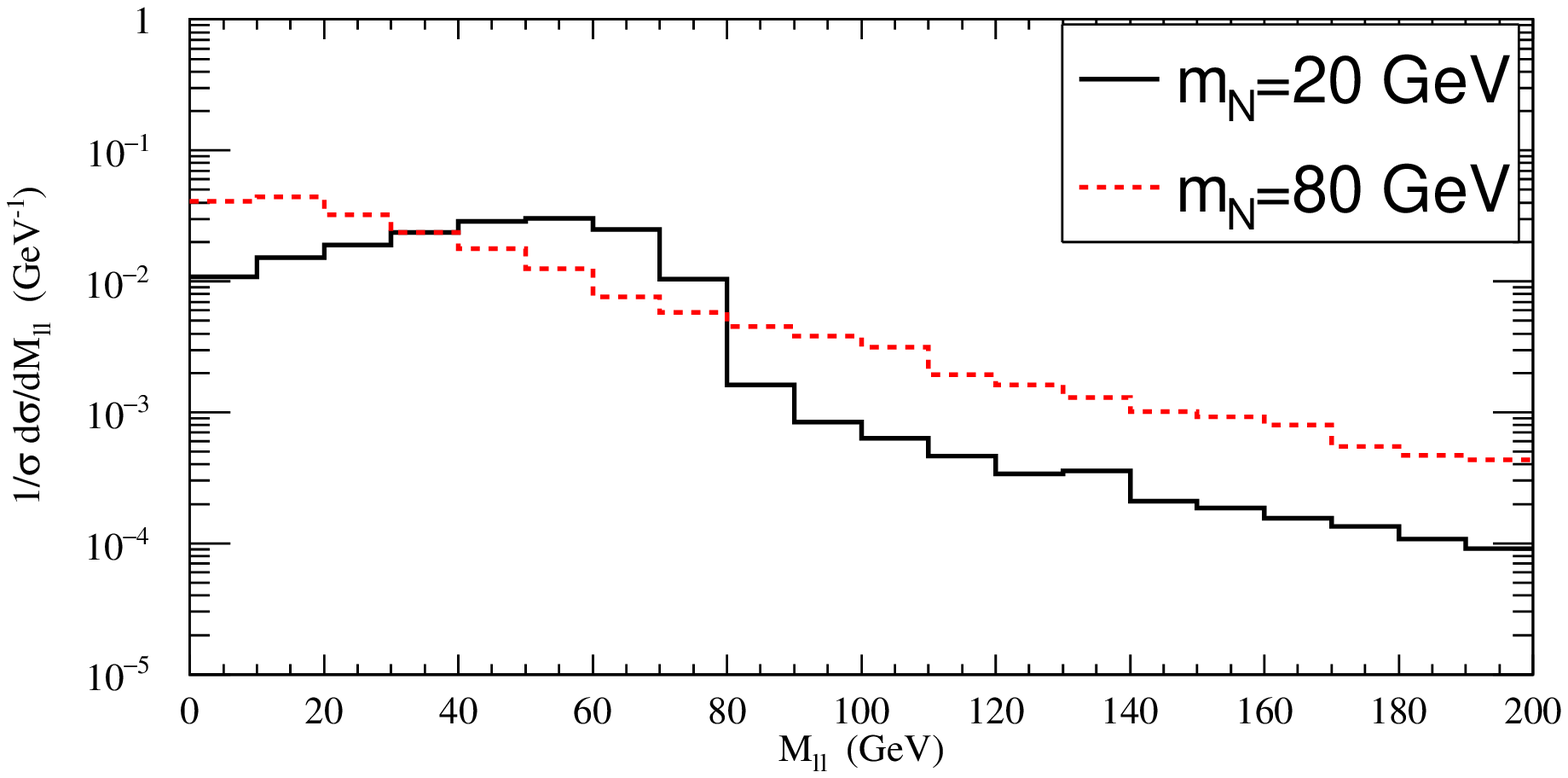} }
\caption{The normalized invariant mass distributions $1/\sigma{\rm d}\sigma/{\rm d}M_{\ell\ell}$ of (a) same-sign dileptons, (b) opposite-sign dileptons for $m_N=20,~80~{\rm GeV}$ at 3.5~TeV FCC-ep.}\label{fig7}
\end{center}
\end{figure}

In the following, taking the process Eq.~(\ref{12}) as an example, we investigate the transverse momentum distributions and reconstructed invariant mass distributions of the final state same-sign dileptons and jets for the process $e^-p \rightarrow e^- \mu^{\pm}\mu^{\pm}+X$.
Our signal process includes 2 same-sign muons and 3 jets besides the electron. The muon (jet) originating from $N$ decay is denoted by $\ell_{\beta}$ ($j_{1,2}$) and that produced in association with $N$ is denoted by $\ell_{\alpha}$ ($j_3$).
For charged-leptons, the transverse momentum differential distribution is defined as $1/\sigma{\rm d}\sigma/{\rm d}p^{\ell}_{\rm T}=1/\sigma({\rm d}\sigma/{\rm d}p^{\ell_\alpha}_{\rm T}+{\rm d}\sigma/{\rm d}p^{\ell_\beta}_{\rm T})/2$. For jets, the transverse momentum of $j_{1,2}$ is found to be softer than $j_3$. Analogously, we define $1/\sigma{\rm d}\sigma/{\rm d}p^{j}_{\rm T}=1/\sigma({\rm d}\sigma/{\rm d}p^{j_1}_{\rm T}+{\rm d}\sigma/{\rm d}p^{j_2}_{\rm T})/2$. In Fig.~\ref{fig5}, we plot the normalized transverse momentum distributions $1/\sigma{\rm d}\sigma/{\rm d}p^{\ell,j,j_3}_{\rm T}$ for $m_N=60~{\rm GeV}$.
The invariant mass of the heavy Majorana neutrino can be reconstructed from the four-momenta of the muon and two soft jets. For the purpose of illustration, we choose FCC-ep with $\sqrt{s}=3.5~{\rm TeV}$. In Fig.~\ref{fig6}, we display the normalized differential distribution $1/\sigma{\rm d}\sigma/{\rm d}M_{\ell jj}=1/\sigma({\rm d}\sigma/{\rm d}M_{\ell_{\alpha}jj}+{\rm d}\sigma/{\rm d}M_{\ell_{\beta} jj})/2$ for the reconstructed invariant mass. The peak positions imply the mass of heavy Majorana neutrino and can be reconstructed effectively. When $m_N$ is much lower than $m_W$, e.g.~$m_N=20~{\rm GeV}$, a second peak appears due to the resonant production of $W$ boson.
We also show the normalized differential distributions for the invariant mass of the lepton pair in Fig.~\ref{fig7}.

\begin{figure}[!htbp]
\begin{center}
\subfigure[]{\label{fig8a}
\includegraphics[width=0.30\textwidth]{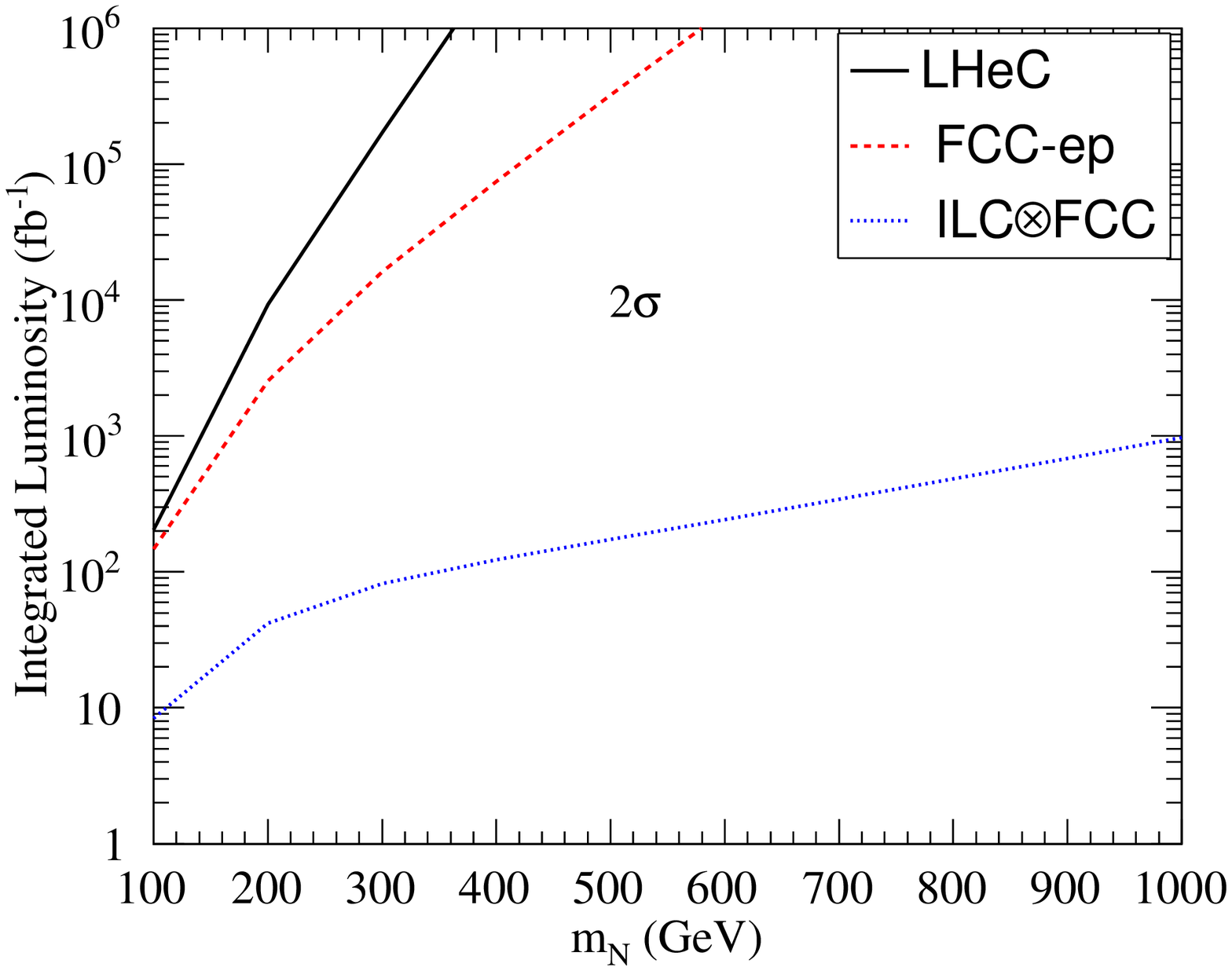} }
\hspace{-0.5cm}~
\subfigure[]{\label{fig8b}
\includegraphics[width=0.30\textwidth]{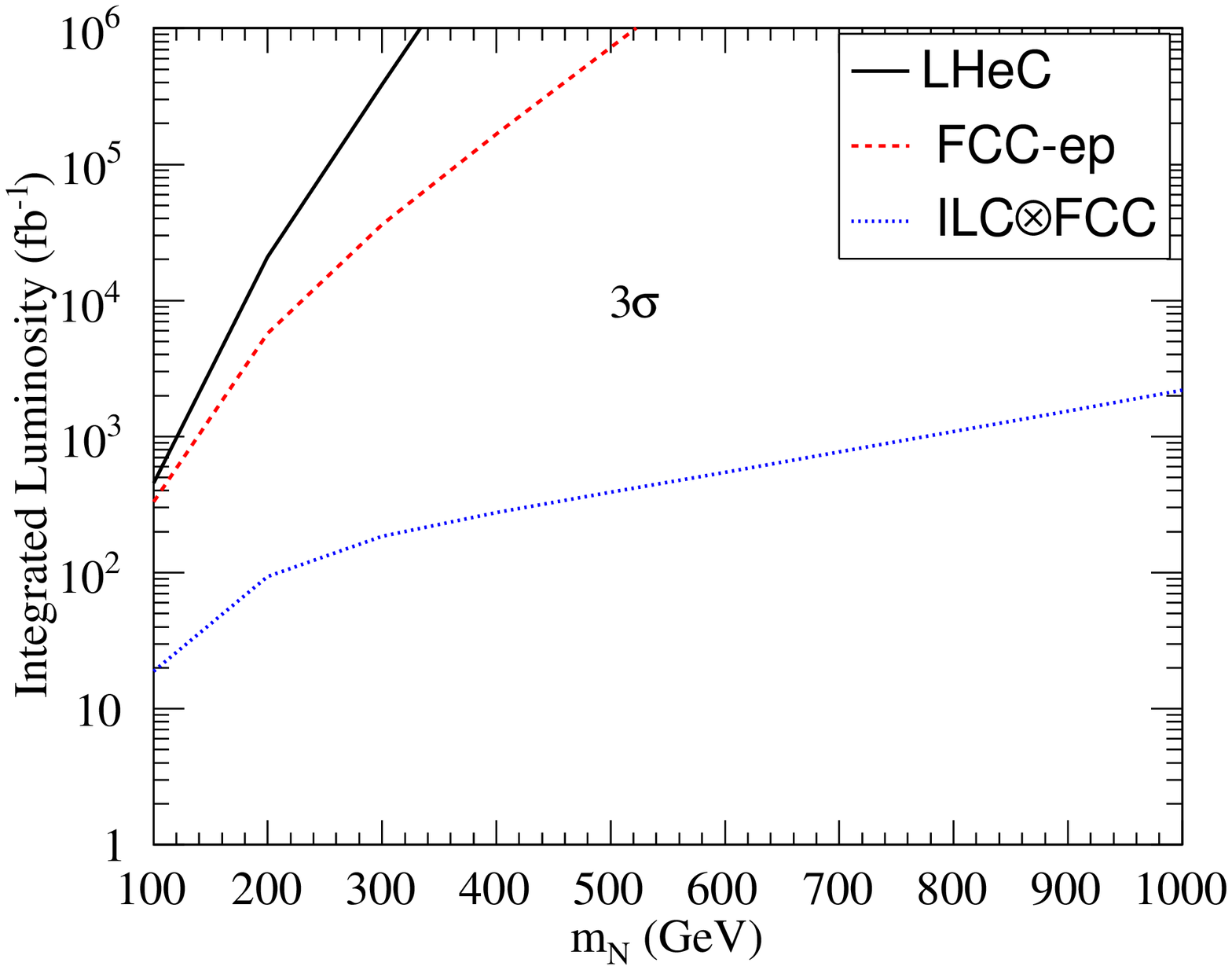} }
\hspace{-0.5cm}~
\subfigure[]{\label{fig8c}
\includegraphics[width=0.30\textwidth]{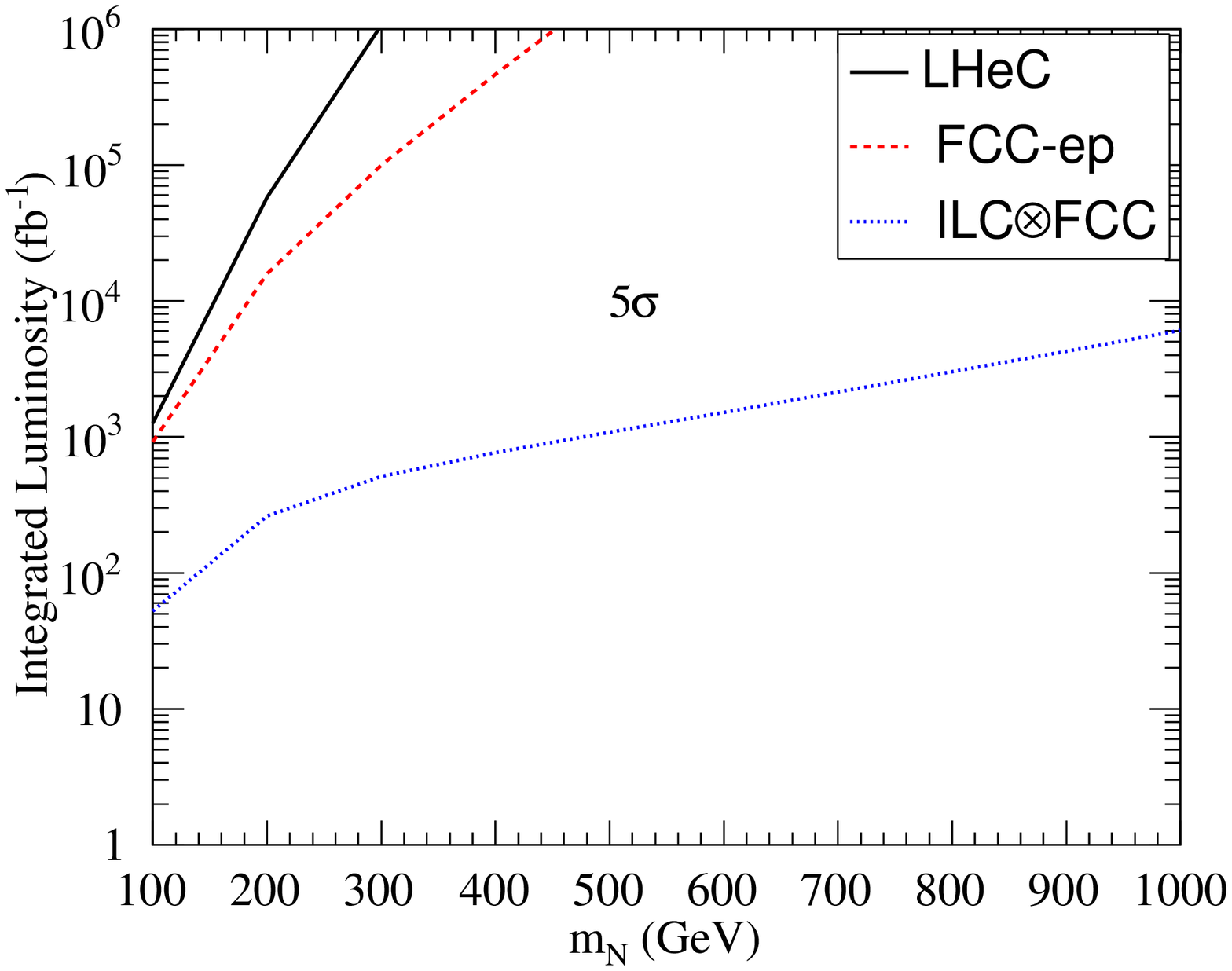} }
\caption{The required integrated luminosity as a function of $m_N$ for (a) 2$\sigma$, (b) 3$\sigma$ and (c) 5$\sigma$ discovery at LHeC (solid), FCC-ep (dash) and ILC$\otimes$FCC (dot).}\label{fig8}
\end{center}
\end{figure}

\begin{figure}[!htbp]
\begin{center}
\includegraphics[width=0.52\textwidth]{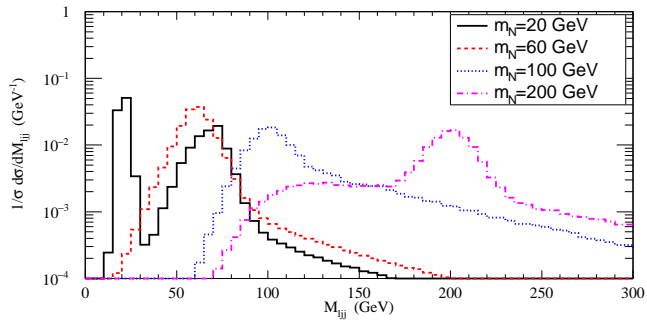}
\caption{The normalized invariant mass distributions $1/\sigma{\rm d}\sigma/{\rm d}M_{\ell jj}$ for the same $m_N$ as Fig.~\ref{fig6} at 3.5~TeV FCC-ep with the cuts.}\label{fig9}
\end{center}
\end{figure}

To be more realistic, we simulate the detector effects by smearing the charged-lepton and jet energies according to the assumption of the Gaussian resolution parametrization
\begin{eqnarray}
\label{20}
\frac{\delta(E)}{E} = \frac{a}{\sqrt{E}}\oplus b,
\end{eqnarray}
where $\delta(E)/E$ is the energy resolution, $a$ is a sampling term, $b$ is a constant term, and $\oplus$ denotes a sum in quadrature. We take $a=5\%$, $b=0.55\%$ for charged-leptons and $a=100\%$, $b=5\%$ for jets, respectively~\cite{Aad:2009wy}.
In order to identify the isolated charged-leptons and jets, we define the angular distribution between particle $i$ and particle $j$ as
\begin{eqnarray}
\label{21}
\Delta R_{ij} = \sqrt{\Delta \phi^2_{ij}+\Delta\eta^2_{ij}}~,
\end{eqnarray}
where $\Delta\phi_{ij}$ ($\Delta \eta_{ij}$) denotes the difference between the particles' azimuthal angle (rapidity).
In the following numerical calculations, we impose the basic acceptance cuts,
\begin{eqnarray}
\label{22}
p_{T}^{\ell} > 10~{\rm GeV} \; ,~~~~~ |\eta^{\ell}| < 2.5 \; ,~~~~~ p_{T}^{j} > 10~{\rm GeV} \; ,~~~~~ |\eta^{j}| < 5 \; ,
\end{eqnarray}
\begin{eqnarray}
\label{23}
\min\{\Delta R_{\ell\ell},\Delta R_{\ell j},\Delta R_{jj}\} > 0.4 \; ,~~~~~ {E\slash}_{T} < 20~{\rm GeV} \; .
\end{eqnarray}
The dominant same-sign di-muon backgrounds in the standard model to our signal process come from the $W$-boson pair production and its leptonical decays:
$e^- p \rightarrow e^-(\nu_e) W^{\pm} W^{\pm} j j j X$ with $W^{\pm} \rightarrow \mu^{\pm} \nu_{\mu}$, which are simulated by MadGraph. Considering the significant impact of the process $e^- p \rightarrow \nu_e \mu^{-} \mu^{-}X$, we also simulate the top quark backgrounds $e^-p \rightarrow \nu_e t\bar{t}W^-X$.
The ``faked backgrounds" come from the detector mis-measurement, e.g.~$3\ell$ (or $4\ell$) + $3j$ final states with one (or two) lepton lost in the beam pipe. We analyse these kinds of backgrounds and find that they are much smaller than the same-sign di-muon backgrounds as expected. These kinds of backgrounds are taken into account in our analysis.
For the muon production channel, the charge mis-identification is negligible due to the almost absence of photons converting to muons~\cite{Aad:2011vj,Chatrchyan:2011wba}.
At 2$\sigma$ (3$\sigma$) [5$\sigma$] significance, the required luminosity as a function of $m_N$ at LHeC, FCC-ep and ILC$\otimes$FCC are displayed in Fig.~\ref{fig8}, where the statistical significance is defined as $S/\sqrt{B}$ with $S$ ($B$) being the signal (background) event numbers after the basic acceptance cuts. It shows that, with the integrated luminosity of $300~{\rm fb}^{-1}$, the heavy Majorana neutrino mass can reach 110 GeV (79 GeV) [74 GeV] at LHeC, 125 GeV (88 GeV) [75 GeV] at FCC-ep and 665 GeV (425 GeV) [220 GeV] at ILC$\otimes$FCC for 2$\sigma$ (3$\sigma$) [5$\sigma$] discovery.
In order to estimate the systematic uncertainty in the background evaluation, we simply introduce a 20\% background uncertainty by scaling the background cross section $\sigma_b$ as $\sigma_b \rightarrow 1.2 \times \sigma_b$, where 2$\sigma$ discovery can be achieved for $m_N=105~{\rm GeV}$ (118 GeV) [610 GeV] at LHeC (FCC-ep) [ILC$\otimes$FCC] with the integrated luminosity of $300~{\rm fb}^{-1}$.
Finally, we investigate the reconstruction of the heavy Majorana neutrino after the cuts in Eqs.~(\ref{22}) and (\ref{23}). It is easy to notice that the heavy Majorana neutrino can be reconstructed from one muon and two soft jets. As an example, we show the normalized differential distribution $1/\sigma{\rm d}\sigma/{\rm d}M_{\ell jj}=1/\sigma({\rm d}\sigma/{\rm d}M_{\ell_{\alpha}jj}+{\rm d}\sigma/{\rm d}M_{\ell_{\beta} jj})/2$ at FCC-ep (3.5 TeV) in Fig.~\ref{fig9}.

\section{Summary}\label{sec4}

The heavy Majorana neutrinos $N_R$ are introduced to explain the minor neutrino mass in the so-called phenomenological type-I seesaw mechanism. Due to the existence of the Majorana neutrino mass term, we can search for the Majorana neutrinos via the lepton-number violating processes, which have been studied in various experiments.
The heavy Majorana neutrino production at $e^{-}p$ colliders is an important complement to that at the hadron collider LHC. The latter is well studied and reviewed in Ref.~\cite{Pascoli:2018heg}, where 2$\sigma$ discovery can be made for $m_N=280~{\rm GeV}$ with the integrated luminosity of $300~{\rm fb}^{-1}$ and $|R_{\mu N}|^2=0.01$ in the charged lepton flavor conservation scenario, while the heavy Majorana neutrino mass can reach 110 GeV (79 GeV) [74 GeV] at LHeC, 125 GeV (88 GeV) [75 GeV] at FCC-ep and 665 GeV (425 GeV) [220 GeV] at ILC$\otimes$FCC for 2$\sigma$ (3$\sigma$) [5$\sigma$] discovery with the integrated luminosity of $300~{\rm fb}^{-1}$.
In this work, we explore the heavy Majorana neutrino production and decay in the context of $W^\ast\gamma$ interaction and  investigate the related dilepton production process at future $e^{-}p$ colliders. The cross sections for the processes $e^{-}p \rightarrow e^- \mu^{\pm}\mu^{\pm}+X$ and $e^{-}p \rightarrow \nu_e \mu^{-}\mu^{\pm}+X$ at future LHeC, FCC-ep and ILC$\otimes$FCC are predicted. We further investigate the process $e^{-}p \rightarrow e^- \mu^{\pm}\mu^{\pm}+X$ in detail, and obtain several differential distributions.
Combined with the results of the heavy Majorana neutrino production via single $W$ exchange, this work can provide helpful information to search for the heavy Majorana neutrinos at future $e^{-}p$ colliders.

\section*{Acknowledgements}

The authors would like to thank Profs. Shun Zhou, Yan-Rui Liu, Hong-Lei Li, Yi Jin and Zhong-Juan Yang for their helpful discussions.
This work is supported in part by National Natural Science Foundation of China (grant Nos. 11325525, 11635009, 11775130, 11875179)
and the Natural Science Foundation of Shandong Province (grant No. ZR2017MA002).

\end{document}